\documentclass[%
groupedaddress,
 amsmath,amssymb,
 aps,
]{revtex4-2}

\usepackage{graphicx}% Include figure files
\usepackage{dcolumn}% Align table columns on decimal point
\usepackage{bm}% bold math
\begin{document}

\preprint{APS/123-QED}

\title{$\mathcal{PT}$-Symmetric Quantum State Discrimination for Attack on BB84 Quantum Key Distribution}

\author{Yaroslav Balytskyi}
\affiliation{University of Colorado Colorado Springs, CO 80918, USA}

\author{Manohar Raavi}
\affiliation{University of Colorado Colorado Springs, CO 80918, USA}

\author{Anatoliy Pinchuk}
\affiliation{University of Colorado Colorado Springs, CO 80918, USA}

\author{Sang-Yoon Chang}
\affiliation{University of Colorado Colorado Springs, CO 80918, USA}

\date{\today}% It is always \today, today,

\begin{abstract}
Quantum Key Distribution or QKD provides symmetric key distribution using the quantum mechanics/channels with new security properties. The security of QKD relies on the difficulty of the quantum state discrimination problem. 
We discover that the recent developments in $\mathcal{PT}$ symmetry can be used to expedite the quantum state discrimination problem and therefore to attack the BB84 QKD scheme. We analyze the security of the BB84 scheme and show that the attack significantly increases the eavesdropping success rate over the previous Hermitian quantum state discrimination approach. We design and analyze the approaches to attack BB84 QKD protocol 
exploiting an extra degree of freedom provided by the $\mathcal{PT}$-symmetric quantum mechanics. 

\end{abstract}

\maketitle

\section{Introduction}\label{intro}
Security and cryptographic mechanisms rely on the secret key between the authorized sender and receiver for the confidentiality or integrity protections. While public-key cryptography offers mechanisms to exchange keys so that the outcome of the public-key exchange yields the symmetric keys, the traditional public-key key exchange based on RSA
and Diffie-Hellman Key Exchange 
are in risk because of the emerging quantum computing. The advancement of quantum computing is both in practice (proof-of-concept quantum computers, e.g., IBM~\cite{IBM} and Google~\cite{Google}) and in the algorithms building on quantum computers (e.g., Shor's algorithm, Ref.~\cite{Shor}). 
While such quantum computing developments can be used for expediting the solving of the problems for beneficial purposes, it can also find its uses for breaking the traditional cryptographic ciphers. The traditional hardness problems anchoring the security of such key exchange algorithms, such as prime factorization problem, can be solved in polynomial time by the attackers equipped with quantum computers. National Institute of Standards Technology (NIST), traditionally influential in standardizing and facilitating the deployment of cryptographic ciphers, e.g., DES, AES, is therefore currently in the multi-year process of standardizing quantum-resistant key exchange ciphers. 

More recent developments for key exchange/distribution use the sender-receiver channels, including those using quantum channels/mechanics, Ref.~\cite{BB84,SP} or wireless signal channels in radio-frequency (RF), Ref.~\cite{BS}, or in electrical field propagation, Ref.~\cite{SY}. 
We focus our study on the emerging Quantum Key Distribution (QKD) which exchanges quantum bits (qubits) between the sender and the receiver over a quantum channel, such as one based on optical communications, Refs.~\cite{WT,Rel1,Rel4}. QKD provides a unique security property that the authorized sender and receiver can detect if the qubit transmissions have been accessed/eavesdropped and the key compromised so that the sender and receiver can distinguish between the secret key vs. the eavesdropped key. 

Unlike the classical public key cryptography, QKD is based on the difficulty of the physical problem of the quantum state discrimination, Ref.~\cite{JB}, and on the no-cloning theorem, Ref.~\cite{WKW}. The goal of the quantum state discrimination is to find in which state the qubit is, and consists of finding an optimal observable and strategy of measurements. The no-cloning theorem ensures that an eavesdropper as able to do the measurement only once, since the qubit cannot be perfectly copied. Since the security of QKD is based on the laws of physics rather than on the hard mathematical problem, it is information-theoretically secure as opposed to assuming a computational bound on the attacker. 

We apply the state of the art research on $\mathcal{PT}$-symmetric quantum mechanics and discover a novel method to increase the eavesdropping success rate of the attacker against BB84 QKD protocol. $\mathcal{PT}$ symmetry enables the attacker to discriminate the quantum states with higher probability, enabling the advanced attack to learn the exchanged keys. We construct three approaches for the attack on BB84 in theory and analyze the practicality and implementation options. 

The rest of the paper is organized as follows. Section~\ref{sec:primer} provides the primer and the background in BB84 QKD protocol as well as a comparison between the regular Hermitian quantum mechanics  vs. the $\mathcal{PT}$ symmetric one. Section~\ref{sec:app} explains how $\mathcal{PT}$ symmetry can advance the quantum state discrimination and, building on that, Section~\ref{Main} describes the three approaches in attacking BB84 and analyzes them. 
Section~\ref{sec:related} discusses the related work focusing on the QKD in practice and the literature studying QKD in information-theoretical security. 
Lastly, Section~\ref{Conclusion} concludes the paper.

\section{A Primer on QKD and $\mathcal{PT}$ Symmetry vs. Hermitian Quantum Mechanics}
\label{sec:primer}

\begin{figure}
  \centering
  \includegraphics[width=.7\linewidth]{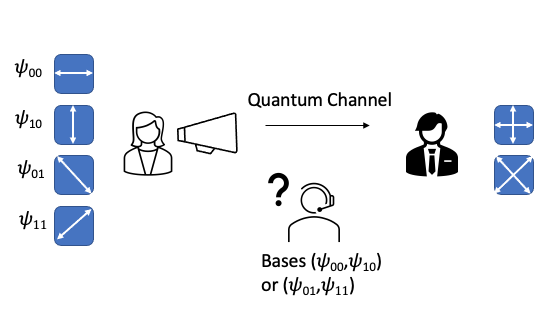}
  \caption{An illustration of the BB84 protocol for the qubit transmission for the key exchange. %The first column corresponds to the qubit sent by Alice, the second one corresponds to the measument applied by Bob on the qubit, and the third one corresponds to the result of that measurement. 
  The sender Alice (left) transmits the qubits in one of the four states, e.g., $\psi_{00}$, while the receiver Bob (right) uses either of the two bases to make measurements on the received qubits.}
  \label{fig:qkd}
\end{figure}

\begin{figure}
  \centering
  \includegraphics[width=.7\linewidth]{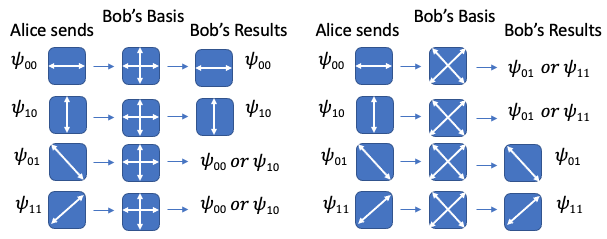}
  \caption{An illustration of four states used in the BB84 protocol and the Bob's receiving/measurement process. The first column shows the sender Alice's state; the second column corresponds to the receiver Bob's basis; and the third column shows the Bob's measurement/decoding result. If Bob's basis matches and aligns with Alice's state, then Bob can correctly decode the qubit with a probability of 1. If Bob's basis does not match/align with Alice's state, the it becomes a coin toss with Bob's decoding between the two states, i.e., the correct qubit with 0.5 probability.  \emph{After} the qubits have been transmitted (not shown in the diagram), the sender shares the bases she used for each of the the qubit transmission so that the receiver and the sender know which qubits were delivered successfully with the matching state/base. These qubits are used to construct the secret key.
  }
  \label{fig:states}
\end{figure}

\subsection{BB84 QKD Protocol}
While in the classical communications the information is encoded in classical bits, in quantum communications, the information is encoded in qubits. This allows to detect the presence of the adversary trying to learn the key since the applied measurement changes the state of the qubit. By comparing the randomly chosen measurements, sender and receiver can detect the presence of an eavesdropper in the channel. 

BB84 QKD protocol, Ref.~\cite{BB84}, uses the set of two bases: 

\begin{equation}
\begin{cases}
|\psi_{00}\rangle = |0\rangle\\
|\psi_{10}\rangle = |1\rangle\\
\end{cases}
 and  \
\begin{cases}\label{Bases}
|\psi_{01}\rangle = |+\rangle = \frac{|0\rangle  + |1\rangle}{\sqrt{2}}\\
|\psi_{11}\rangle = |-\rangle = \frac{|0\rangle  - |1\rangle}{\sqrt{2}}\\
\end{cases}  \\
\end{equation}

Note, the conversion between the first and second basis can done by the Hadamard gate:

$$
\mathcal{H}  = \frac{1}{\sqrt{2}}\begin{pmatrix}
 1 & 1\\
 1 & -1
\end{pmatrix}
$$

Alice generates two threads of random classical bits, $a$ and $b$, of equal length $l$ and encodes them using the bases in Eqn.~\ref{Bases}, by the following block of qubits:

$$
|\psi\rangle = \bigotimes_{i = 1}^{l} |\psi_{a_ib_i}\rangle
$$

which is sent through an open channel to Bob, as illustrated in Fig.~\ref{fig:qkd}. Bob generates the thread of random classical bits $c$ of the same length $l$ and makes a measurement in the basis specified by $c$.
Then, Alice and Bob through an open classical channel establish in which cases the classical bits from $b$ and $c$ coincide using the results of the measurements in these cases as the shared secret key and discard the rest.

To demonstrate an advantage of the $\mathcal{PT}$-symmetric quantum mechanics over the Hermitian one for an attack on BB84, we consider the simplest Eve's attack in both approaches.

In the conventional Hermitian quantum mechanics, Eve who is trying to learn the key guesses the basis correctly in $50\%$ of the cases, and in the other $50\%$ of the cases the results of her measurements becomes a coin toss with a random output, Fig.\ref{fig:states}. Therefore, on average, Eve is able to guess $75\%$ of the bits correctly sent by Alice to Bob. 

The attacker on BB84 has to discriminate between the four possible states which are used for the encryption and this is a difficult problem in the Hermitian quantum mechanics. Even in the case of two non-orthogonal quantum states, it is not possible to discriminate between them by a single measurement. 

BB84 and QKD in general rely on both the quantum state discrimination problem and the no-cloning theorem. Our research focuses on a  $\mathcal{PT}$-symmetric quantum state discrimination to enhance the eavesdropping rate.

\subsection{$\mathcal{PT}$ Symmetry vs. Hermitian}
$\mathcal{PT}$-symmetric quantum mechanics is a complex extension of the regular Hermitian quantum mechanics, Refs.~\cite{CMBDCB1,CMBDCB2,AM}, and provides additional opportunities to solve the quantum state discrimination problem. In this approach, the condition of Hermiticity of the Hamiltonian, $\mathcal{H} = \mathcal{H}^{\dagger}$, is replaced by a more general requirement of $\mathcal{PT}$ symmetry. The Hamiltonian $\mathcal{H}$ is defined as $\mathcal{PT}$- symmetric if it satisfies the requirement $\mathcal{H} = \mathcal{H}^{\mathcal{PT}}$.

The action of the parity operator $\mathcal{P}$  changes the sign of the quantum-mechanical
coordinate $\hat{x}$ and the momentum $\hat{p}$ operators:

$$
\mathcal{P}\hat{x}\mathcal{P} = - \hat{x}; \ \ \mathcal{P}\hat{p}\mathcal{P} = - \hat{p}
$$

Up to the unitary transformation, $\mathcal{P}$ operator is defined as: 

$$
\mathcal{P} =
\begin{pmatrix}
0 & 1 \\
1 & 0
\end{pmatrix}  
$$

The anti-linear time-reversal operator $\mathcal{T}$ flips the signs of $\hat{p}$ and the imaginary unit $i$ while leaving $\hat{x}$ invariant:

$$
\mathcal{T}\hat{x}\mathcal{T} = \hat{x}; \ \ \mathcal{T}\hat{p}\mathcal{T} = - \hat{p}; \ \  \mathcal{T}i\mathcal{T} = - i
$$

In comparison with the Hermitian case, a remarkable property of $\mathcal{PT}$-symmetric Hamiltonian is that in addition to governing the time evolution of the quantum state, it allows to manipulate an inner product of two states providing an extra degree of freedom. 

Firstly, it can be exploited for speeding up the quantum evolution. After being theoretically predicted, Ref.~\cite{FasterT}, this effect was demonstrated experimentally Ref.~\cite{FasterO}. 
Secondly, this additional degree of freedom plays a crucial role for the quantum state discrimination problem enabling, \textit{in principle}, single-measurement two-state quantum state discrimination, Ref.~\cite{Disc2}.

However, such an approach involves a similarity transformation of the initial Hilbert space in which the state vectors are defined into a space spanned by the eigenvectors of the $\mathcal{PT}$-symmetric Hamiltonian. This transformation can be only done with the efficiency less than one, and the measurement may produce a null result. Consequently, even though \textit{in principle} $\mathcal{PT}$-symmetric quantum state discrimination involves a single measurement to discriminate $N = 2$ states, it can provide inconclusive results (provides definite results with a probability less than one) even in the absence of the noise. 
This feature of $\mathcal{PT}$-symmetric quantum discrimination resembles an unambiguous state discrimination developed in the framework of the regular Hermitian quantum mechanics,
Ref.~\cite{Ivanovic}.

$\mathcal{PT}$-symmetric devices for quantum state discrimination are currently under experimental investigation, Ref.~\cite{ObsDist}, promising practical implementations in the near future. In Section~\ref{sec:related}, we provide an estimate on the efficiency $\eta$ which has to be achieved in such devices in order to be of practical relevance for an attack on BB84.

\vspace{0.1in}\hspace{-0.2in}
\textbf{Our Contributions.}\indent Using the novel $\mathcal{PT}$-symmetric quantum state discrimination, Ref.~\cite{Disc2}, we show how to achieve an enhancement of the eavesdropping rate for an attack on BB84 by approximately $10.5\%$ in comparison with the regular Hermitian case.

\section{Using $\mathcal{PT}$ Symmetry for the Quantum State Discrimination Problem}\label{sec:app} 
In this section, we apply $\mathcal{PT}$ symmetry for quantum state discrimination problem which in turn can be used for attacking BB84. The Hamiltonian satisfying the requirement of $\mathcal{PT}$-symmetry has the following general form, Ref.~\cite{Disc2}: 
\begin{equation}\label{Hamiltonian}   
H = H^{\mathcal{PT}} = 
\begin{pmatrix}
r e^{i\theta} & s \\
s & r e^{-i\theta}
\end{pmatrix}
\end{equation}

where $r$, $s$ and $\theta$ are real parameters. The $\alpha$ parameter defined by the parameters of $\mathcal{PT}$-symmetric Hamiltonian, Eqn.(\ref{Hamiltonian}), $\sin\left(\alpha\right) = \frac{r}{s}\sin\left(\theta\right)$ manifests an additional degree of freedom provided by $\mathcal{PT}$-symmetry. Variation of the $\alpha$ allows to manipulate an inner product of two quantum states. 

In comparison with the Hermitian quantum mechanics, $\mathcal{PT}$-symmetric quantum mechanics has an additional $\mathcal{C}$ operator which depends on the $\alpha$ parameter:
\begin{equation}\label{C_op}
\mathcal{C} =\frac{1}{\cos\left(\alpha\right)}
\begin{pmatrix}
i \sin\left( \alpha \right) & 1 \\
1 & -i \sin\left( \alpha \right)
\end{pmatrix}  
\end{equation}

In the limit $\alpha\rightarrow0$ this operator coincides with the regular $\mathcal{P}$ operator. The $\mathcal{CPT}$ scalar product of two vectors
$|\lambda\rangle$ and $|\mu\rangle$, in the $\mathcal{PT}$-symmetric quantum mechanics, is defined as:

\begin{equation}\label{CPT_product}
\langle\lambda|\mu\rangle = \left(\mathcal{CPT} \lambda\right)^T\cdot\mu
\end{equation}

where $T$ refers to the transposition of a matrix.

Unlike the Hermitian case where the scalar product is fixed, variation of the matrix elements in $\mathcal{PT}$-symmetric Hamiltonian, Eqn.(\ref{Hamiltonian}), transforms non-orthogonal vectors into orthogonal ones, Ref.~\cite{FasterT}. 

In the case when the number of states is $N = 2$, $\mathcal{PT}$-symmetric quantum mechanics provides two alternative solutions for the state discrimination, Ref.~\cite{Disc2}:
\begin{itemize}
\item \textit{Solution 1: Adjusting $\mathcal{PT}$-symmetric Hamiltonian in order to make two quantum states orthogonal under the $\mathcal{CPT}$ scalar product.}
\item \textit{Solution 2:
Evolving two quantum states by the $\mathcal{PT}$-symmetric Hamiltonian to make them orthogonal by a regular Hermitian scalar product.}
\end{itemize}

\textit{Solution 1} uses the variation of $\alpha$ parameter to adjust the $\mathcal{CPT}$ scalar product, Eqn.~\ref{CPT_product} while \textit{Solution 2} takes advantage of the fact that $H \ne H^\dagger$ and effectively the Hermitian scalar product is changed by the following matrix, Ref.~\cite{Disc2}:

$$
\cos^2\left(\alpha\right)e^{iH^\dagger t}e^{-iH t} = $$
\begin{equation}\label{Evolved}
\begin{pmatrix}
\cos^2\left(\omega t - \alpha\right) + \sin^2\left(\omega t\right) & -2i\sin^2\left(\omega t\right)\sin\left(\alpha\right) \\
2i\sin^2\left(\omega t\right)\sin\left(\alpha\right) & \cos^2\left(\omega t + \alpha\right) + \sin^2\left(\omega t\right)
\end{pmatrix}  
\end{equation}

Note, in the limit $\alpha\rightarrow 0$ it reduces to the unit matrix and coincides with the regular Hermitian case. In our solution described in the Section~\ref{Main}, we consider attacks on BB84 exploiting both these solutions and different parts of the parameter space of the $\mathcal{PT}$-symmetric Hamiltonian, Eqn.~\ref{Hamiltonian}.

\section{Attack on BB84}\label{Main}
In this section, we propose three alternative attack approaches and show that with the use of $\mathcal{PT}$-symmetric quantum mechanics it is possible to correctly guess the fraction $\frac{5\eta}{6}$ of the encoded bits sent by Alice to Bob in comparison with $\frac{3}{4}$ in a regular Hermitian case, where $\eta$ is the efficiency of the similarity transformation mentioned in Section~\ref{intro}. Although all these three approaches give the same efficiency for an attacker in theory, they have different implications when implementing them in practice. 

\subsection{Approach 1: Unambiguous exclusion of one of the states}
First, we consider an option which allows to unambiguously exclude one of the states, Ref.~\cite{Bal}. 

In $\mathcal{PT}$-symmetric quantum mechanics the Hilbert space can effectively by curved similarly to black hole curving the space in general relativity, Ref.~\cite{FasterT}. This makes the positions on the Bloch sphere in-equivalent. First, application of the following gate:
$$
R_1 = 
\begin{pmatrix}
1 & 0\\
0 & i
\end{pmatrix}
$$

allows to convert our states to a form convenient for the subsequent $\mathcal{CPT}$ measurement:

$
\begin{cases}
|\psi_{00}\rangle \rightarrow |0\rangle\\
|\psi_{10}\rangle \rightarrow |1\rangle\\
\end{cases}
$ and $
\begin{cases}
|\psi_{01}\rangle \rightarrow \frac{|0\rangle  + i|1\rangle}{\sqrt{2}}\\
|\psi_{11}\rangle \rightarrow  \frac{|0\rangle  - i|1\rangle}{\sqrt{2}}\\
\end{cases} \\
$

Now, observe that the $\mathcal{CPT}$ scalar product of the transformed states $|\psi_{01}\rangle$  and $|\psi_{11}\rangle$
vanishes for an arbitrary value of the $\alpha$ parameter: 

$$\left(\langle\psi_{01}|\psi_{11}\rangle \right)_{\mathcal{CPT}} = 0
$$

since 

$$
\left(\langle\psi_{01}|\right)_{\mathcal{CPT}} = \frac{\left(1 + \sin\left(\alpha\right)\right)}{\sqrt{2}\cos\left(\alpha\right)}
\begin{pmatrix}
1 \\
-i
\end{pmatrix}^T
$$

This allows to build the following $\mathcal{CPT}$ projection operators:
$$
P_{1}^1 = \left(\frac{|\psi_{01}\rangle\langle\psi_{01}|}{\langle\psi_{01}|\psi_{01}\rangle}\right)_{\mathcal{CPT}}=  \frac{1}{{2}}
\begin{pmatrix}
1 & -i\\
i & 1
\end{pmatrix}
$$

$$
P_{2}^1 = \left(\frac{|\psi_{11}\rangle\langle\psi_{11}|}{\langle\psi_{11}|\psi_{11}\rangle}\right)_{\mathcal{CPT}}=  \frac{1}{{2}}
\begin{pmatrix}
1 & i\\
-i & 1
\end{pmatrix}
$$

which are the $\mathcal{CPT}$ observables since:
 
$$
\left[\mathcal{CPT}, P^1_{1, 2}\right] = 0
$$

and the corresponding $\mathcal{CPT}$ measurement:

$$
\hat{\mathcal{M}_1} = P_1^1 - P_2^1
$$

The corresponding cosines of angles between our transformed states take the form:
$$
\cos^2\left(|\psi_{01}\rangle, |\psi_{00}\rangle\right) = \cos^2\left(|\psi_{01}\rangle, |\psi_{10}\rangle\right) = \frac{1 + \sin\left(\alpha\right)}{2}
$$

$$
\cos^2\left(|\psi_{11}\rangle, |\psi_{00}\rangle\right) = \cos^2\left(|\psi_{11}\rangle, |\psi_{10}\rangle\right) = \frac{1 - \sin\left(\alpha\right)}{2}
$$

$$
\cos^2\left(|\psi_{01}\rangle, |\psi_{11}\rangle \right) = 0
$$
Taking the limit $\alpha\rightarrow\pm\frac{\pi}{2}$, one of two states $|\psi_{01}\rangle$ or
$|\psi_{11}\rangle$ can be eliminated depending on the $\pm$ sign chosen. For example, if we take the limit $\alpha\rightarrow \frac{\pi}{2}$, the measurement produces the result $\mathcal{M}_1 = - 1$ only in the case when the state is $|\psi_{11}\rangle$ and the result $\mathcal{M}_1 = 1$ means that the qubit is in one of the three states, $|\psi_{00}\rangle$, $|\psi_{10}\rangle$ or $|\psi_{01}\rangle$. 

Therefore, an application of this approach would provide an attacker an unambiguous knowledge of an encoded state in $25\%$ of the cases. On average, if Eve guesses the base wrongly which happens in $50\%$ of the time, she can correctly guess the state of the qubit in fraction of $\frac{2}{3}$ of the cases in comparison with $\frac{1}{2}$ in the Hermitian case. Therefore, on average, Eve correctly guesses the encoded bit in $\frac{5\eta}{6}$ of the cases. 

However, such an approach involves taking the limit $\alpha\rightarrow\pm\frac{\pi}{2}$  meaning that the absolute values of the matrix elements of the $\mathcal{PT}$-symmetric Hamiltonian have to be large, and exactly at $\alpha = \pm \frac{\pi}{2}$ the $\mathcal{C}$ operator, Eqn.~\ref{C_op}, and metrics, Eqn.~\ref{CPT_product}, become singular (the so-called $\mathcal{PT}$-symmetry breaking point).

As a result, this approach may be challenging for the practical implementation. Therefore, we consider an alternative solution in the next subsection which involves  moderate values of the $\alpha$ parameter and the $\mathcal{CPT}$ measurements.

\subsection{Approach 2: $\mathcal{CPT}$ measurement with $\alpha\not\to\pm\frac{\pi}{2}$}

We show that it is possible to achieve the same average result without taking the limit $\alpha\rightarrow\pm\frac{\pi}{2}$. To exploit the fact of that the Hilbert space is curved, we apply the following gate:
\begin{equation}\label{Gate}
R_2 = 
\begin{pmatrix}
\cos\left(\frac{\rho}{2}\right) & i\sin\left(\frac{\rho}{2}\right)\\
 i\sin\left(\frac{\rho}{2}\right) & \cos\left(\frac{\rho}{2}\right)
\end{pmatrix}
\end{equation}

and adjust the value of $\rho$ to put our states in the convenient positions for the subsequent $\mathcal{CPT}$ measurement.

After application of this gate, the angles between our reference states in terms of the $\mathcal{CPT}$ scalar product become:

$$
\cos\left(|\psi_{00}\rangle, |\psi_{10}\rangle\right) = \frac{\sin\left(\alpha\right)\sin\left(\rho\right)}{\sqrt{1 - \cos^2\left(\rho\right)\sin^2\left(\alpha\right)}}
$$

$$
\cos\left(|\psi_{00}\rangle, |\psi_{01}\rangle\right) = \frac{1 + \sin\left(\alpha\right)\left(\sin\left(\rho\right) + \cos\left(\rho\right)\right)}{\sqrt{2\left(1 + \cos\left(\rho\right)\sin\left(\alpha\right)\right)\left(1 + \sin\left(\rho\right)\sin\left(\alpha\right)\right)}}
$$

$$
\cos\left(|\psi_{00}\rangle, |\psi_{11}\rangle\right) = \frac{1 + \sin\left(\alpha\right)\left(\cos\left(\rho\right) - \sin\left(\rho\right)\right)}{\sqrt{2\left(1 + \cos\left(\rho\right)\sin\left(\alpha\right)\right)\left(1 - \sin\left(\rho\right)\sin\left(\alpha\right)\right)}}
$$

$$
\cos\left(|\psi_{10}\rangle, |\psi_{11}\rangle\right) = \frac{1 - \sin\left(\alpha\right)\left(\sin\left(\rho\right) + \cos\left(\rho\right)\right)}{\sqrt{2\left(1 - \cos\left(\rho\right)\sin\left(\alpha\right)\right)\left(1 - \sin\left(\rho\right)\sin\left(\alpha\right)\right)}}
$$
 
$$
\cos\left(|\psi_{01}\rangle, |\psi_{11}\rangle\right) = \frac{\sin\left(\alpha\right)\cos\left(\rho\right)}{\sqrt{1 - \sin^2\left(\rho\right)\sin^2\left(\alpha\right)}}
$$
$$
\cos\left(|\psi_{10}\rangle, |\psi_{01}\rangle\right) = \frac{1 + \sin\left(\alpha\right)\left(\sin\left(\rho\right) - \cos\left(\rho\right)\right)}{\sqrt{2\left(1 - \cos\left(\rho\right)\sin\left(\alpha\right)\right)\left(1 + \sin\left(\rho\right)\sin\left(\alpha\right)\right)}}
$$

$$
\cos\left(|\psi_{10}\rangle, |\psi_{11}\rangle\right) = \frac{1 - \sin\left(\alpha\right)\left(\sin\left(\rho\right) + \cos\left(\rho\right)\right)}{\sqrt{2\left(1 - \cos\left(\rho\right)\sin\left(\alpha\right)\right)\left(1 - \sin\left(\rho\right)\sin\left(\alpha\right)\right)}}
$$
 
$$
\cos\left(|\psi_{01}\rangle, |\psi_{11}\rangle\right) = \frac{\sin\left(\alpha\right)\cos\left(\rho\right)}{\sqrt{1 - \sin^2\left(\rho\right)\sin^2\left(\alpha\right)}}
$$

Plugging $\alpha = \frac{\pi}{4}$ and $\rho = \frac{3\pi}{4}$ makes $\cos\left(|\psi_{00}\rangle, |\psi_{11}\rangle\right) = 0$. This allows to build the following projection operators:

$$
P_{1}^2 
=\left(\frac{|\psi_{00}\rangle\langle\psi_{00}|}{\langle\psi_{00}|\psi_{00}\rangle}\right)_{\mathcal{CPT}}= 
\frac{1}{{2}}
\begin{pmatrix}
1 - \frac{\sin\left(\rho\right)}{1 + \cos\left(\rho\right)\sin\left(\alpha\right)} & \frac{-i\left(\cos\left(\rho\right) + \sin\left(\alpha\right) + \cos\left(\rho\right)\sin\left(\alpha\right)\right)}{1 + \cos\left(\rho\right)\sin\left(\alpha\right)}\\
\frac{i\left(\cos\left(\rho\right) + \sin\left(\alpha\right) + \cos\left(\rho\right)\sin\left(\alpha\right)\right)}{1 + \cos\left(\rho\right)\sin\left(\alpha\right)} & 1 + \frac{\sin\left(\rho\right)}{1 + \cos\left(\rho\right)\sin\left(\alpha\right)}
\end{pmatrix}
$$

$$
P_{2}^2 
=\left(\frac{|\psi_{11}\rangle\langle\psi_{11}|}{\langle\psi_{11}|\psi_{11}\rangle}\right)_{\mathcal{CPT}}= 
\frac{1}{{2}}
\begin{pmatrix}
1 + \frac{\sin\left(\rho\right)}{1 + \cos\left(\rho\right)\sin\left(\alpha\right)} & \frac{i\left(\cos\left(\rho\right) + \sin\left(\alpha\right) + \cos\left(\rho\right)\sin\left(\alpha\right)\right)}{1 + \cos\left(\rho\right)\sin\left(\alpha\right)}\\
\frac{-i\left(\cos\left(\rho\right) + \sin\left(\alpha\right) + \cos\left(\rho\right)\sin\left(\alpha\right)\right)}{1 + \cos\left(\rho\right)\sin\left(\alpha\right)} & 1 - \frac{\sin\left(\rho\right)}{1 + \cos\left(\rho\right)\sin\left(\alpha\right)}
\end{pmatrix}
$$

which are also the $\mathcal{CPT}$ observables:

$$
\left[\mathcal{CPT}, P^2_{1, 2}\right] = 0
$$

Then, applying the following measurement:
$$
\hat{\mathcal{M}_2} = P_1^2 - P_2^2
$$

and identifying the result of the measurement $\mathcal{M}_2$ with the value of the encoded bit as:
$$
\begin{cases}
\mathcal{M}_2 = 1 \Rightarrow  a = 0\\
\mathcal{M}_2 = -1 \Rightarrow a = 1
\end{cases}
$$

allows to correctly guess the encoded bit in $\frac{5\eta}{6}$ fraction of the cases.

However, one potential disadvantage of such an approach is that in practice it may be simpler to implement the Hermitian measurements instead of the $\mathcal{CPT}$ one. Therefore, we consider one additional option which involves Hermitian measurements instead.

\subsection{Approach 3: Evolution and the Hermitian measurements}

We use an alternative \textit{Solution 2} of the $\mathcal{PT}$-symmetric quantum state discrimination problem involving the non-Hermitian evolution resulting in an effective change of the Hermitian scalar product, Eqn.~\ref{Evolved}. 

First, we apply the same gate as we used before, Eqn.~\ref{Gate}, which puts two of our states in the convenient positions for the further Hamiltonian evolution with $\sigma = \frac{\pi}{4}$:

$$
|\psi_{00}\rangle \rightarrow \begin{pmatrix}
\cos\left(\frac{\pi - 2\sigma}{4}\right)  \\
-i\sin\left(\frac{\pi - 2\sigma}{4}\right)
\end{pmatrix};  \ |\psi_{11}\rangle \rightarrow \begin{pmatrix}
\cos\left(\frac{\pi + 2\sigma}{4}\right)  \\
-i\sin\left(\frac{\pi + 2\sigma}{4}\right)
\end{pmatrix}
$$

The Hamiltonian evolution is performed for a time $\tau$ given by the equation:

\begin{equation}\label{time}
\sin^2\left(\omega \tau\right) = \frac{\cos^2\alpha \cos\sigma}{2\sin\alpha - 2\sin^2\alpha\cos\sigma}
\end{equation}

As a result, $\langle\psi_{00}|\psi_{11}\rangle_{Hermitian} = 0$ and analogously to the \textit{Approach 2} this allows to couple the two bases. 

After the time given by an Eqn.~\ref{time} our four states are converted to: 

$$
|\psi_{a_i, b_i}\rangle \rightarrow e^{-iH\tau}|\psi_{a_i, b_i}\rangle
$$

where the evolution operator is given by:

$$ 
e^{-iH\tau} = \frac{e^{-i r\cos\left(\theta\right)\tau}}{\cos\left(\alpha\right)}
\begin{pmatrix}
\cos\left(\omega \tau - \alpha\right) & -i\sin\left(\omega \tau \right) \\
-i\sin\left(\omega \tau \right) & \cos\left(\omega \tau + \alpha\right)
\end{pmatrix}  
$$

The resulting Hermitian projection operators are:
\begin{equation}\label{Appr3}
P_{1, 2}^3 
=\left(\frac{|\psi_{00, 11}\rangle\langle\psi_{00, 11}|}{\langle\psi_{00, 11}|\psi_{00, 11}\rangle}\right)_{Hermitian}
\end{equation}

Analogously, the measurement is constructed as 
$$
\mathcal{M}_3 = P_{1}^3 - P_{2}^3  
$$

And we identify the result of the measurement $\mathcal{M}_3 = 1$ as $a_i = 0$ and $\mathcal{M}_3 = -1$ as $a_i = 1$.

Note, the $\mathcal{PT}$-symmetric Hamiltonian evolution preserves the $\mathcal{CPT}$ norm of the states since $\left[\mathcal{C}, H\right] = 0$ but changes the Hermitian norm of the states. Therefore, we normalize our states in the projection operators in Eqn.~\ref{Appr3} correspondingly.

\begin{figure}
  \centering
  \includegraphics[width=9cm]{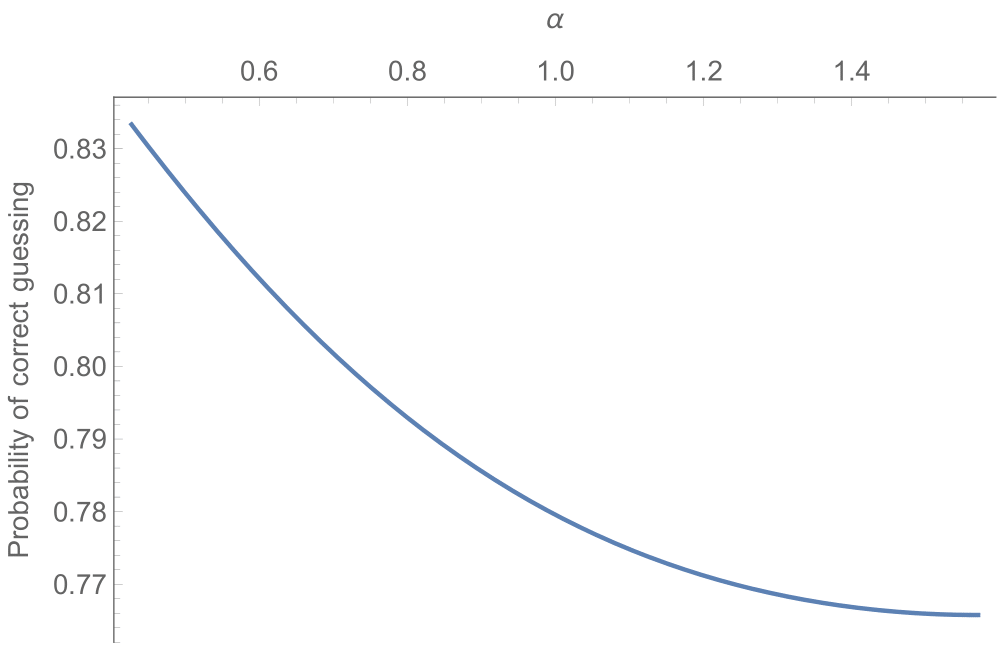}
  \caption{The probability of correct guessing of the encoded bit as a function of the $\alpha$ parameter. For $\alpha < \tan ^{-1}\left(\sqrt{\frac{1}{2} \left(\sqrt{2}-1\right)}\right)$, solution does not exist.}
  \label{Correctness_Hermitian}
\end{figure}

Now we need to find an optimal value of parameter $\alpha$ in such a way that it minimizes the average error in determining which bit was encoded by Alice.
Performing an analogous calculations to the \textit{Approach 2} and expressing the cosines between the evolved states which now are governed by Eqn.~\ref{Evolved}, we plot the corresponding average probability of correct guessing in Fig.~\ref{Correctness_Hermitian}. Note, there is a minimal value of the parameter $\alpha$ which makes such Hamiltonian evolution possible such that $\sin^2\left(\omega\tau\right) \le 1$ in Eqn.~\ref{time}. Additionally, note that this minimal value of $\alpha$ corresponds to the biggest possible value of probability of correct guessing in this approach and which is the same as in the previously considered  \textit{Approaches 1} and \textit{2} and equals $\frac{5\eta}{6}$. This optimal value equals:

$$
\alpha_{Optimal} = \tan ^{-1}\left(\sqrt{\frac{1}{2} \left(\sqrt{2}-1\right)}\right) \approx 0.427079 
$$

\subsection{Comparison Between Three Approaches}

All three approaches give the same \textit{average} probability of the eavesdropper to correctly guess the encoded bit of $\frac{5\eta}{6}$. However, the \textit{Approach 1} has an advantage that in $25\%$ of the cases allows to \textit{unambiguously} determine the state which is used for encoding the bit.

However, \textit{Approach 1} requires the parameters of the $\mathcal{PT}$-symmetric Hamiltonian, Eqn.~\ref{Hamiltonian}, to be close to the the $\mathcal{PT}$-symmetry breaking point which may be challenging for the practical implementation.

\textit{Approach 2} involves moderate values of the $\alpha$ parameter and should be therefore simpler for the practical implementation. 

However, currently many of the $\mathcal{PT}$-symmetric Hamiltonians are implemented as on optical devices where the Hermitian measurements are simpler to implement, Refs.
~\cite{Opt1,Opt2,Opt3,Opt4}. Therefore, the \textit{Approach 3} uses an intermediate values of the $\alpha$ parameter far from the the $\mathcal{PT}$-symmetry breaking point as well as the Hermitian measurements.

Finally, we estimate the required efficiency of such devices in order for these \textit{Approaches} to be relevant in practice. The efficiency of the $\mathcal{PT}$-symmetric quantum state discriminator has to satisfy the requirement:
$$
\frac{5p}{6} > \frac{3}{4} \Rightarrow \eta > 90\%.
$$

The most recent experimental implementation on the $\mathcal{PT}$-symmetric quantum discrimination, Ref.~\cite{ObsDist}, provides a practical experimental platform for the future studies but does not investigate the efficiency for this purpose.

\section{Related Work}
\label{sec:related}
In this section, we discuss the related work in QKD and, more specifically, its practical relevance in implementations and how it offers information-theoretic security.

\subsection{QKD in Practice}
Over the past few years, QKD has gained attention for its unique security benefits. Continuous research to find the practical relevance yielded in multiple simulations, implementations, and modeling frameworks, Refs. ~\cite{Rel1,Rel2,Rel3}. Decoy-state QKD uses attenuated coherent light sources in place of perfect single photon sources and is used in the experiments conducted in Refs.~\cite{Rel1,Rel2}. It helps detection of photon-splitting eavesdropping and can be used in high loss channels. The authors in Ref.~\cite{Rel2} demonstrated the first experimental implementation of decoy-state QKD over telecom fiber. The experimentation include implementations of one-decoy protocol over 15km telecom fiber and weak+vacuum protocol over 60km of telecom fiber.
The authors in Ref.~\cite{Rel1} demonstrated the possibility of global-scale quantum networks by successfully performing decoy-state QKD, using BB84, between a ground observation station and a satellite at an altitude of around 500KM. 
The empirical experiments in these work highlight the unique security benefits by QKD.  
Furthermore, the first measurement-device-independent (MDI) QKD realization in the free-space  over the 19.2km in atmosphere, Ref.~\cite{Rel4}, has been demonstrated approaching a practical satellite-based QKD.

Other work facilitate the practical implementations by providing a framework building on QKD security relying on the use of a single-photon source and fundamental laws of security. Device imperfections and practical implementation limitations play a huge role in keeping QKD security intact. The authors in Ref. \cite{Rel3} proposed a modeling framework that helps as a reference for BB84 QKD. The proposed framework identifies the device imperfections, engineering limitations, and design trade-offs. 

Our approaches can be used to enhance an eavesdropping rate in these and similar devices. Therefore, while constructing the QKD in practice, one has to keep in mind these possibilities if an efficient $\mathcal{PT}$-symmetric device for quantum state discrimination is implemented on practice.

\subsection{QKD for Information-Theoretic Security}

According to Shannon, Ref.~\cite{Sh}, the system is defined to achieve \emph{perfect secrecy} if the mutual information between the ciphertext and the plaintext is zero: $I\left(\mathcal{C}, \mathcal{M}\right) = 0$. As a result, 
the amount of entropy in the key must be greater or equal than that in the message, $H\left(\mathcal{K}\right)\ge H\left(\mathcal{M}\right)$. \emph{One-time pad} encryption by the Vernam cipher, Ref.~\cite{Vernam}, fulfills this condition and is provably secure encryption scheme. However, for its execution, it requires a large key to be exchanged between the communicating parties, which prohibits its deployment in many computing/networking applications. 

QKD provides a unique opportunity to  meet this challenge and make Vernam cipher of practical relevance. For example, in Ref.~\cite{Rel4} more than $3.5\times10^6$ sifted keys  were exchanged over the distance of approximately 20 km through the atmosphere in 13.4 hours. Nevertheless, QKD can also provide the keys for any other symmetric-key cryptosystems. 

Furthermore, the no-cloning theorem preserves the information-theoretic security. The no-cloning theorem on which the security of the QKD rely, Ref.~\cite{WKW}, is based on the linearity of quantum mechanics and applies both to the Hermitian as well as the $\mathcal{PT}$-symmetric quantum mechanics since they differ by the symmetry properties of the Hamiltonian and are both linear. However, as we show in this paper, the eavesdropping efficiency in the $\mathcal{PT}$-symmetric quantum mechanics can be higher in comparison with the Hermitian one if the corresponding device is implemented with sufficient efficiency.

\section{Conclusion}\label{Conclusion}
Recent advances in quantum mechanics and the $\mathcal{PT}$- symmetric  quantum state discrimination can be exploited for attacking and eavesdropping on the key transmissions in the BB84 QKD protocol. We construct three approaches for attacking BB84, all of which have the same performances for guessing the secret bit at $\frac{5\eta}{6}$.  

While our contributions and validations are theoretical, we analyze the practicality and the relevance of our work. The choice between the \textit{Approach 1, 2} or \textit{3} depends on the accessibility of the $\mathcal{CPT}$ and Hermitian measurements as well as the availability of the parameter space of the $\mathcal{PT}$-symmetric  Hamiltonian, Eqn.~\ref{Hamiltonian}. 
As QKD becomes more viable and popular with generating/sharing the keys for securing the next-generation systems, our work informs the security of QKD and facilitates greater research in the direction. We also call for greater research bridging physics/quantum mechanics and cybersecurity as we prepare for the quantum computing era.

\end{document}